\documentclass[preprint,showpacs,preprintnumbers,amsmath,amssymb]{revtex4}

\usepackage{graphicx}
\usepackage{dcolumn}
\usepackage{bm}

\def\be{\begin{equation}}
\begin{document} 


\title{Energy and Angular Momentum Densities of\\ Stationary Gravity Fields}

\author{D. Lynden-Bell}
  \affiliation{Institute of Astronomy, The Observatories, Cambridge CB3 0HA, U.K.}
  \altaffiliation{And Institute of Theoretical Physics,  Charles University, 180 00 Prague 8, Czech Republic.}
\author{Joseph Katz}
   \affiliation{The Racah Institute of Physics, Givat Ram, 91904 Jerusalem,
Israel\,}
  \altaffiliation{Also Institute of Astronomy, The Observatories, Cambridge CB3 0HA, 
U.K. and Institute of Theoretical Physics, Charles University, 180 00 Prague 8, Czech Republic.} 

\author{Ji\u{r}\'{i} Bi\v{c}\'{a}k}
  \affiliation{Institute of Theoretical Physics,  Charles University, 180 00 Prague 8, Czech Republic\,}
  \altaffiliation{And Institute of Astronomy, The Observatories, Cambridge CB3 0HA, 
U.K.}

\date{\today}

\begin{abstract}
We give physical explanations of explicit invariant expressions for
the energy and angular momentum densities of gravitational fields in
stationary space-times. These expressions involve non-locally defined
conformal factors. In certain coordinates these become locally defined
in terms of the metric. These results are derived via expressions for total gravitational potential energy from the difference between the total energy and the mechanical energy. The latter involves kinetic energy seen in the frame of static observers.

When in the axially symmetric case we consider zero angular momentum
observers (who move orthogonally to surfaces of constant time), we find
that the angular momentum they attribute to the gravitational field is
solely due to their motion. 
\end{abstract}
  
\pacs{04.20  -q.\,\,\, 04.20.Cv}
\maketitle

\section{\label{sec:intro}Introduction}

The aim of this paper is to show that in the special case of stationary asymptotically flat spacetimes there are invariantly defined expressions for both the energy density of the gravitational field and for the angular momentum density. Furthermore, these expressions are the gravitational analogues of $(E^2+B^2)/(8\pi)$ and $[\vec r\times (\vec E\times\vec B)]/(4\pi)$ for electromagnetism. However the gravitational expressions do involve the gradiant of a conformal factor which is defined non-locally using a geometrical structure depending on the Killing vector. We hope that our expressions can be generalized to slowly varying systems in general relativity. We do not aim to introduce any new generally defined concept of energy or quasi-local energy such as those of Penrose, Hawking or Hayward [1]. We are concerned to isolate the gravitational potential energy in the stationary case and reexpress it in terms of `fields'  whereas those quasi-local expressions contain material energy as well as gravitational energy and go over to the ADM mass at infinity.

Misner, Thorne and Wheeler (1973) [2], hereafter MTW, deny the existence
of localised gravitational field energy-density (with somewhat
strident rhetoric). Nevertheless they give an expression for the
localised gravitational energy in the exceptional case of spherical
symmetry (quoting Misner \& Sharp (1964) [3] to include time
dependence). Katz (2005) [4] has recently given an expression for it
in the more general case of conformastationary metrics. As our
expression (which agrees with his) differs from that of MTW even for
the spherically symmetric case, we shall start by considering their
special case in this introduction.

Landau \& Lifshitz (1966) [5] show that the general spherically
symmetric metric can be put in the form (setting $c^2=1$ and writing
$\xi^2$ for their $e^\nu$)
$$
ds^2=\xi^2dt^2-\left(\frac{dr^2}{1-\frac{2m(r,t)}{r}}+r^2d\hat{\bf
r}^2\right)~,
$$
where $m(r,t)$ tends to $m_\infty=GM_\infty$ the total mass at infinity times $G$, and
$$
\hat{\bf r}=(\sin\theta\cos\varphi , \sin\theta\sin\varphi ,
\cos\theta)~,~(d\hat{\bf r})^2 =d\theta^2 +\sin^2\theta d\varphi^2~.
$$

The relevant component of Einstein's equations reads

\begin{equation} \label{eq:1}
\frac{\partial m}{\partial r}=4\pi G r^2 T^0_0~,~~{\rm
so}~~~m(r,t)=m_\infty-\int_r^\infty 4\pi Gr^2 T^0_0(r,t)dr
\end{equation}
$T^0_0$ is the energy density of the matter in the rest frame
$(r,\theta,\phi)$ constant. When $r=2m(r,t)$ at some radius $r_*(t)$ a central black hole of that radius and mass $M_*(t)=r_*/2G=m_*/G$ is present. Such a case is considered in [6] but here we shall confine ourselves to global complete spacetimes.

To see exactly what equation (\ref{eq:1}) means consider for
example a fluid of dust. Then $T^{\mu\nu}=\rho u^\mu u^\nu$ where
$\rho$ is the density of the fluid in its rest frame and $u^\mu$ is
its four velocity which is of the form $(u^0,u^r,0,0)$ by symmetry.

Let $w^\mu$ be the 4-velocity of static observers with $(r,\theta,\phi)$ constants, so these
are unit vectors and in their coordinates have only one component
$w^\mu=\left( \xi^{-1},0, 0,0\right),~w_\mu=(\xi,0,0,0)$. Since the
fluid moves radially in our frame, they see it to have a greater density
due to Lorentz contraction. This increased density of rest mass is
$\rho u_\mu w^\mu\equiv \rho {\bf u.w}$ and the \emph{mechanical energy}
density in any frame now including kinetic energy is $\frac{\rho({\bf
u.w})}{\sqrt{1-v^2/c^2}}$. This is the time component of 
\begin{equation} \label{eq:2}
\rho{\bf u.w}u^\nu =T^\nu_\mu w^\mu,
\end{equation}
which is the flow vector of mechanical energy and momentum seen by the
static observers. The mechanical energy is its flux through a
space-like hypersurface such as $t=$ constant. The part of that flux
through $t=$ constant that lies within $r=$ constant will be 
\be  \label{eq:3}
E_M(r)=\int^r_0 T^0_\mu w^\mu\sqrt{-g}~d^3x=\int^r_0T^0_0 4\pi
r^2\left(1-\frac{2m(r,t)}{r}\right)^{-1/2}dr. 
\end{equation}
$E_M(r)$ is a special case of the energy seen by observers moving orthogonally to a hypersurface like that geometrically defined by Wald [12]. 
Although for illustrative clarity we considered only dust above, this
is still the mechanical energy within  $r$ whatever constitutes the
$T_{\mu\nu}$. Indeed for a perfect fluid it includes the internal
energy as well as rest mass and kinetic energy,
\begin{equation} \label{eq:4}
T^0_0=\frac{\rho+pv^2}{ 1-v^2}~,
\end{equation}
where $v$ is the 3-velocity of the fluid in their frame.

A comparison between equations (\ref{eq:1}) and (\ref{eq:3})
illustrates that (\ref{eq:1}) is seductively like the classical
relationship between density and mass but in fact conceals all the
complications beneath a cloak of apparent clarity.

Firstly $T^0_0$ is not the density in these coordinates unless $v$
is zero; secondly $4\pi r^2 dr$ is not the volume element which is
rather 
\begin{equation} \label{eq:5}
4\pi r^2 [1-2m(r,t)/r]^{-1/2}dr=dV~.
\end{equation}

Outside the matter $m$ becomes constant and gives $G$ times the total
mass-energy seen from infinity. MTW, quoting Misner \& Sharp [2], argue
that $M(r,t)=G^{-1}m(r,t)$ is the total mass-energy within $r$ at time $t$ and is
only less than $E_M(r,t)$ because of $E_G(r,t)$ the gravitational
energy which is negative. Indeed, since $E_M$ (equation (\ref{eq:3}))
contains the rest-mass energy, the kinetic energy and the internal
energy within the matter, the difference

\begin{equation} \label{eq:6}
E_M(r,t)-G^{-1}m(r,t)=\int^r_0\left\{\left[1-\frac{2m(r,t)}{r}\right]^{-1/2}
-1\right\}T^0_0 4\pi r^2 dr=-E_G(r,t) 
\end{equation}
must be due to gravitation.

This would lead to a gravitational energy density of 

$$
\frac{dE_G(r)}{dV}=-\left[1-\left(1-\frac{2m(r,t)}{r}\right)^{1/2}\right]
T^0_0=-\left[1-\left(1-\frac{2m(r,t)}{r}\right)^{1/2}\right]
\left(\frac{1}{4\pi Gr^2}\right) \frac{\partial m}{\partial r}~~,
$$
which is only non-zero inside matter.

To see that this deduction might not be water-tight, we turn to the
electrostatic analogue in flat space. The electrical energy of a
spherical charge distribution, $Q(r)$ being within $r$, can be
calculated by starting at the centre and imagining the distribution to
be built up by adding shells of charge consecutively. The shell with
charge $dQ$ is added when the potential at $r$ is $Q(r)/r$ so the
electrical energy up to $r$ is 
$$
E_{em}(r)=\int^r_0\frac{Q(r)}{r} \frac{dQ}{dr} dr.
$$
As further shells are added we merely extend the upper bound of the
integral so the part within $r$ does not change. However it would be
wrong to deduce that the energy density in the electric field at $r$
in flat space is
$$\frac{1}{4\pi r^2} \frac{dE_{em}(r)}{dr}= \frac{Q}{4\pi r^3}
\frac{dQ}{dr}~.
$$
An alternative is found from the formula
$E^*_{em}(r)=\int^r_0\frac{1}{2}\phi  (dQ/dr)dr$ where $\phi$ is
the electrical potential $\phi=\int^\infty_r (Q/r^2)dr$. This
formula takes account of the fact that the electrical potential
changes from $Q/r$ as outer shells are added. However it is also wrong
to imagine that the electrical energy density is $(4\pi
r^2)^{-1}(dE_{em}^*/dr) =(8\pi r^2)^{-1} \phi
(dQ/dr)$. Notice that both of the above formulae only give
contributions from within the charge distribution $Q(r)$ whereas the
true answer due to Maxwell is
$(8\pi)^{-1}\int E^2 dV=(8\pi)^{-1}\int (Q/r^2)^2 dV$. Of course all three expressions integrate over space to
the same total electrical energy, what is in dispute is the {\it
distribution} of that energy. This electrical analogy suggests that it
might be appropriate to evaluate (\ref{eq:6}) to infinity and
integrate the result by parts. Then it might be possible to express
the result as the volume integral of a perfect square which might be
interpreted as in the electrical case.

Doing this we write $x=2m(r,t)/r$ and use (\ref{eq:1}) 
\begin{eqnarray*}
-GE_G&=&\int^\infty_0[(1-x)^{-1/2}-1] \partial m/\partial
r~dr=\frac{1}{2} \int^\infty_0[(1-x)^{-1/2}-1]\left(x+r\frac{\partial
x}{\partial r}\right)dr\\
&=&\frac{1}{2} \int^\infty_0\left\{[(1-x)^{-1/2}-1]
x+r\frac{\partial}{\partial r} [2-2(1-x)^{1/2}-x]\right\}dr~.
\end{eqnarray*}
Integrating by parts and remembering that $x \rightarrow 0$ as $r
\rightarrow \infty$ the two terms recombine to give us just such a
perfect square integrated over the volume $dV$   given by (\ref{eq:5})
viz
\begin{equation} \label{eq:7}
-E_G= \int \frac{1}{8\pi Gr^2} \left[
 1-\left(1-\frac{2m(r,t)}{r}\right)^{1/2}   \right]^2dV=\frac{1}{8\pi G}\int F^2 dV~.
\end{equation}
where we have introduced a gravitational field strength $F$ given by
$$
F=\frac{1}{r}
\left[1-\left(1-\frac{2m}{r}\right)^{1/2}\right]
=-(-g_{rr})^{-1/2}\frac{\partial\Phi}{ \partial r}~,
$$
where $\Phi$ is a gravitational potential. Equation (\ref{eq:7}) may be seen as the gravitational analogue of Maxwell's $\int E^2dV/(8\pi)$
 but in the special case of spherical symmetry. For the geometrical interpretation of $\Phi$ see below. An expression equivalent to (\ref{eq:7}), but expressed in isotropic coordinates was given in [4]. 
 Denoting by $\rightarrow$ values at large $r$,
$$
F_r=-\partial\Phi/\partial r =\frac{1}{r}
\left[\left(1-\frac{2m}{r}\right)^{-1/2}-1\right] \rightarrow \frac{m}{r^2}~.
$$
Evidently
\begin{equation} \label{eq:8}
\Phi=\int^\infty_r\frac{1}{r}
\left\{\left[1-\frac{2m(r,t)}{r}\right]^{-1/2}-1\right\}dr \rightarrow
\frac{m}{r}~.
\end{equation}
 
 What physical meaning should be ascribed to this gravitational field
and what is the source of $\Phi$?

$$\nabla^2\Phi=\frac{1}{r^2}\left(1-\frac{2m}{r}\right)^{1/2}
\frac{\partial}{\partial
r}\left[\left(1-\frac{2m}{r}\right)^{1/2}r^2\frac{\partial\Phi}{\partial
r}\right] =-\frac{1}{r^2} \frac{\partial m}{\partial r}
+\frac{1}{r^2}\left[\sqrt{1-\frac{2m}{r}}-\left(1-\frac{m}{r}\right)\right] ~,
$$
so
\begin{equation} \label{eq:9}
-\nabla^2\Phi=   \frac{1}{2} \left(\kappa T^0_0- F^2 \right)
 = \frac{1}{2} \left[\kappa T^0_0-  (\nabla\Phi)^2\right],
\end{equation}
where $\kappa=8\pi G$. This shows that the negative gravitational field energy acts alongside
the positive $T^0_0$ as a source for $\Phi$. In this last form it is
natural to generalise (\ref{eq:9}) out of spherical symmetry and this
is indeed what   has already been done in  [4] for conformastatic and
conformastationary systems. 

We now turn to the interpretation of $\Phi$. All spherical spaces have
conformally flat spatial metrics. If we rewrite our spatial metric in
isotropic form,
$$
d\sigma^2= e^{2\Lambda}(d\overline{r}^2+\overline{r}^2d\hat{\bf r}^2)~,
$$
then, at constant $t$,
$$
 \left( 1-\frac{2m(r,t)}{r}\right)^{-1/2}dr=e^\Lambda d\overline{r}
$$
and
$$
r=e^\Lambda\overline{r}~,
$$
so
$$
\left[\left(1-\frac{2m}{r}\right)^{-1/2}-1\right] \frac{dr}{r} =
d\ln\overline{r} -d\ln r=d\ln \left(\frac{\overline{r}}{r}\right)~.
$$
Hence, taking $\frac{\overline{r}}{r}\rightarrow 1$ at infinity, we
find, cf equation (\ref{eq:8}),
\begin{equation} \label{eq:10}
e^{-\Lambda}=\frac{\overline{r}}{r}=\exp \left\{-\int^\infty_r
\left[\left(1-\frac{2m}{r}\right)^{-1/2}-1\right] \frac{dr}{r}\right\} =e^{-\Phi}~.
\end{equation}
So $\Phi$ is $\Lambda$ and $e^{2\Phi}$ is $-g_{11}$ in isotropic
coordinates. For time dependent spherical metrics $r$ is geometrically defined as a real radius and the surfaces of constant $t$ are geometrically defined as those symmetric cuts of spacetime orthogonal to $r$.

If we make the conformal transformation $d\overline{\sigma}^2=e^{-2\Phi}d\sigma^2$,
then
\begin{equation} \label{eq:11}
d\overline{\sigma}^2 = (d\overline{r}^2+\overline{r}^2
d\hat{\bf r}^2). 
\end{equation}

Thus the interpretation of $\Phi$ for the spherically symmetric case
is that $e^{-2\Phi}$ is the conformal factor of the transformation
which makes the spatial metric flat. In general, spatial metrics of
general spaces cannot be made flat, most of those that can have
already been explored in [4]. It is worth noticing that
the energy (\ref{eq:7}) or (\ref{eq:3}) depends solely on the spatial
metric, $\xi$ is not involved.

Although we can not generally make a conformal transformation to give
a flat 3-space, we can in practice make a conformal transformation to a
3-space which has no \emph{scalar} 3-curvature. This turns out to be a
crucial step in determining the invariant energy density of general
stationary gravitational fields.

In Katz, Lynden-Bell \& Bi\v{c}\'{a}k (2006) [6] this technique is
employed to give energy densities according to both static and
hypersurface orthogonal observers, such as ZAMOs (Bardeen (1970) [7]).

Here we give a more physical exposition of that work and extend it to
include angular momentum density. This leads us to a new and different
expression for gravitational field energy $\mathcal{V}$.

For stationary space times we write our metric in the alternative
forms
\begin{equation} \label{eq:12}
\left.
\begin{array}{ll}
ds^2=\xi^2(dt-{\mathcal{A}}_k dx^k)^2-\gamma_{kl}dx^kdx^l=g_{\mu\nu}dx^{\mu}dx^\nu\\
\hspace{7mm}=\zeta^2 dt^2-\tilde{\gamma}_{kl}(dx^k-W^kdt)(dx^l-W^ldt)
\end{array}\\ \right\}
\end{equation}
where Greek indices run from 0 to 3 and latin ones from 1 to
3. Evidently
\begin{equation} \label{eq:13}
\mathcal{A}_k=-g_{0k}/g_{00}~~~{\rm and
}~~~\xi^2\mathcal{A}_k=-g_{0k}=-\tilde{\gamma}_{kl}W^l=-\xi_k
\end{equation}
where $\xi^\mu$ is the `stationary' Killing vector.
We define 
$$
\gamma={\rm det}(\gamma_{kl}) 
$$
and the equivalent for $\tilde\gamma$.

Landau and Lifshitz [5]  show that for a given spacetime there are many stationary metrics of the form  (\ref{eq:12}) since an arbitrary function of the $x^k$ can be added to $t$ without destroying the stationary form of the metric. Thus there are many different slicings of the given spacetime into time and space. However, as discussed in [6],  we can get a unique geometrically defined slicing by demanding that it be maximal in the sense  that the trace of the the external curvature, $K$, of the constant time slices be zero. This choice of time slicing is clearly a good one in that it picks out Boyer-Lindquist time in Kerr spacetime. We shall  thereafter make this choice of the time coordinate so that the space on each constant time slice is a well defined geometrical concept.

\section{Gravitational Field Energy Densities}

In [4] it was realised that the definition of mechanical energy used in
spherical symmetry by MTW [1] could be extended to any stationary
space-time. If $w^\mu=\xi^\mu/\xi$, then the density of rest mass in a
dust fluid seen by a static observer is
$\frac{\rho}{\sqrt{1-v^2}}=\rho({\bf{u.w}})$ and the energy density on any
spacelike surface element $d\Sigma_\mu$ is $\rho({\bf
u.w})u^\mu\sqrt{-g}d\Sigma_\mu=T^{\mu\nu}w_\nu\sqrt{-g}d\Sigma_\mu ~$\footnote{Here 
$d\Sigma_\mu=\frac{1}{3!}\epsilon_{\mu\nu\rho\sigma} dx^\nu \wedge   dx^\rho\wedge  dx^\sigma$, $\epsilon_{\mu\nu\rho\sigma}$ is the permutation symbol in 4 dimensions    with
$\epsilon_{0123}=1$ and in 3 dimensions it is $\epsilon_{klm}$ with $\epsilon_{123}=1$.}. If
we have a more complicated $T_{\mu\nu}$ such as a gas or a plasma this
last expression still measures the mechanical energy density as seen
by the static observers but it now includes internal energy (including
rest mass energy) and the kinetic energy of motion relative to the
observers. If we sum these local contributions we get the total
mechanical energy with no contribution from the gravitational binding
energy
\begin{equation} \label{eq:14}
E_M=\int T^\mu_\nu w^\nu~\sqrt{-g}d\Sigma_\mu =\int T^0_0~dV,~~~{\rm where}~~~dV=\sqrt{\gamma}d^3x.
\end{equation}
We notice that this expression agrees with that given for spherical
symmetry (\ref{eq:3}) and (\ref{eq:6}). However, this is not generally the same as Wald's expression because $\xi^\mu$ and $w^\mu$ are not generally hypersurface orthogonal, so are not normal to the hypersurface. When black holes are absent we may now define the total
gravitational energy of any stationary space-time by $E_G=M-E_M$,
where $M$ is the total mass. We get an interesting expression
for $E_G$ by using Einstein's equations to transform $T^0_0$ into
`field' quantities via integrations by parts analogously to Maxwell's
treatment in electrodynamics.

We write ${\bf \nabla}$ for the vector operator in the 3-space with
the $\gamma_{kl}$ metric and then follow Lynden-Bell \&
Nouri-Zonoz (1998) [8] (see also Nat\'ario (2000) [9]) in writing
$\mathcal{B}^k=\eta^{ijk} \partial_j A_k$, where $\eta^{ijk}=(\sqrt{\gamma})^{-1}\epsilon^{ijk}$. The 4-vector
$\mathcal{B}^\lambda$ is {\it invariantly} defined by
$\mathcal{B}^\lambda
=-\eta^{\lambda\mu\nu\sigma}\partial_\mu(\xi_\nu/\xi^2)\xi_\sigma\xi
=(0,\mathcal{B}^k)$. Then
$\mathcal{B}^2=\frac{1}{2}\partial_j\mathcal{A}_k \nabla^{[j}
\mathcal{A}^{k]}$ where square brackets around indices denote the
anti-symmetric part. Landau \& Lifshitz [5] denote
$\partial_i\mathcal{A}_j-\partial_j \mathcal{A}_i$ by $f_{ij}$ so,
translating their expression of Einstein's equations to our notation and writing $\mathcal{E}=-\nabla\ln \xi$,
\begin{equation}\label{eq:15}
\xi^2\left(\xi^{-1}\nabla^2\xi +\frac{1}{2}\xi^2\mathcal{B}^2\right)
=R_{00},
\end{equation}
\begin{equation} \label{eq:16}
-\xi^{-1}\nabla^k\nabla^l\xi+\frac{1}{2}\xi^2(\gamma^{kl}\mathcal{B}^2
-\mathcal{B}^k \mathcal{B}^l) + P^{kl}=R^{kl},
\end{equation}
\begin{equation} \label{eq:17}
-\frac{1}{2}\xi^2[\nabla\times{\bf{\mathcal{B}}} -3 \mathcal{E} \times
 {\bf{\mathcal{B}}}]^k=R^k_0, 
\end{equation}
where $P^{kl}$ is the curvature tensor of the $\gamma_{kl}$ 3-space
formed from $\gamma_{kl}$ just as $R^{\mu\nu}$ is formed from
$g_{\mu\nu}$ in 4-space. Now
$R=g_{\mu\nu}R^{\mu\nu}=(w_\mu w_\nu-\gamma_{\mu\nu})
R^{\mu\nu}= \xi^{-2}R_{00}-\gamma_{kl}R^{kl}$, where
$\gamma_{\mu\nu}=w_\mu w_\nu -g_{\mu\nu}$ is a 4-dimensional covariant
version of $\gamma_{kl}$. From (\ref{eq:15}) and (\ref{eq:16})
\begin{equation} \label{eq:18}
2\xi^{-1}\nabla^2\xi-\frac{1}{2} \xi^2 \mathcal{B}^2 - P = R,
\end{equation}
from (\ref{eq:15}) and (\ref{eq:18})
\begin{equation} \label{eq:19}
\kappa T_{00}=R_{00} -\frac{1}{2} \xi^2 R =\xi^2\left(\frac{3}{4}\xi^2
\mathcal{B}^2 +\frac{1}{2} P\right)~.
\end{equation}
Now in the introduction we found that an important step in making the
field energy a perfect square was the introduction of a conformal
transformation which yielded a flat 3-space. We can not do that
generally but we can transform so that our new 3-space has a vanishing
scalar curvature. We write
$\overline{\gamma}_{kl}=e^{-2\Phi}\gamma_{kl}$ and use the
relationships between the curvature of two conformally related spaces
given, e.g., in Stephani et al. (2003) [10], equation 3.85 (contracted)
$$
e^{-2\Phi}\overline{P}=P+4\nabla^2\Phi -2(\nabla\Phi)^2~.
$$
So the transformation that makes $\overline{P}$ zero obeys 
\begin{equation} \label{eq:20}
-2\nabla^2\Phi+(\nabla\Phi)^2=\frac{1}{2}P,
\end{equation}
which we may rewrite like a Schrodinger equation by setting
$\overline{\psi} =e^{-\frac{1}{2}\Phi}$ so that
\begin{equation} \label{eq:21}
\nabla^2\overline{\psi}- \frac{1}{8}P\overline{\psi}=0~.
\end{equation}
Equation (\ref{eq:20}) is the generalisation of equation (\ref{eq:9})
of the introduction and using (\ref{eq:19}) takes the form
\begin{equation}
-\nabla^2\Phi=\frac{1}{2}\left[\xi^{-2}\kappa T_{00}
 -\frac{3}{4 }\xi^2 \mathcal{B}^2
- |\nabla\Phi|^2\right],
\end{equation}
which shows how both gravomagnetic and gravitational field strengths
subtract from the `source' of $\Phi$. However previously that source
was written in terms of $\kappa T^0_0$ not $\kappa T_{00}$ and since
our expression (\ref{eq:14}) for mechanical energy also involves
$\kappa T^0_0$ we now rewrite our basic equations
(\ref{eq:15})--(\ref{eq:18}) in terms of $\kappa T^0_0$, $\kappa
T^k_0$ and $\kappa T^{kl}$. To do this we use equations (\ref{eq:19})
and (\ref{eq:17}) and aim to get only second derivative terms on the
left $\kappa T^0_0=\xi^{-2}\kappa T_{00} +A_k R^k_0$, so
\begin{equation} \label{eq:23}
\frac{3}{4}\xi^2\mathcal{B}^2+\frac{1}{2} P-\frac{1}{2}\xi^2
 \mathcal{A}\cdot\left(\nabla\times \mathcal{B}-3 \mathcal{E} \times
  \mathcal{B} \right) =\kappa T^0_0~.
\end{equation}
Using (\ref{eq:20}) for $P$ and
$-\frac{1}{2}\xi^2{\bf{\mathcal{A}}}\cdot\nabla\times{\bf{\mathcal{B}}}=\frac{1}{2}
\nabla\cdot(\xi^2 {\bf{\mathcal{A}}} \times {\bf{\mathcal{B}}} )
-\frac{1}{2}\mathcal{B}\cdot\nabla\times (\xi^2{\bf{\mathcal{A}}})$, we find on
simplifying the right-hand side of (\ref{eq:24})
\begin{equation} \label{eq:24}
\nabla\cdot\left(\frac{1}{2}\xi^2{\bf{\mathcal{A}}} \times
{\bf{\mathcal{B}}} -2{\bf\nabla}\Phi\right) \equiv \nabla\cdot\mathcal{D}=\kappa(T^0_0+\rho_G),
\end{equation}
where
\begin{equation} \label{eq:25}
\kappa \rho_G =-\frac{1}{4}\xi^2{\mathcal{B}}^2-|\nabla\Phi|^2-\frac{1}{2}\xi^2
{\mathcal{A}}\cdot(  \mathcal{E} \times  \mathcal{B})~;
\end{equation}
and from (\ref{eq:17})

\begin{equation} \label{eq:27}
(\nabla\times{\mathcal{B}})^k
=-2\kappa\xi^{-2}T^k_0+3(\mathcal{E}\times  \mathcal{B})^k~.
\end{equation}
Notice that all quantities on the right of  (\ref{eq:25}) are defined in terms of a Killing vector and  (the square of the gradiant of) a conformal factor which, being the solution of the `invariant' elliptic equation (\ref{eq:20}), depends only on the slicing, its geometry and on the boundary conditions.

Equations (\ref{eq:24}) and (26) may be thought of as a
non-linear generalisation of Maxwell's equations and indeed we shall
see presently that the terms of (\ref{eq:25}) constitute the energy
density of the gravitational field.

The final equation of this trio comes from (\ref{eq:16}) and
(\ref{eq:18}) and reads
\begin{equation} \label{eq:28}
\xi^{-1}(\gamma^{kl}\nabla^2\xi-\nabla^k\nabla^l\xi)=\kappa T^{kl}
-\left(P^{kl}- \frac{1}{2}\gamma^{kl}P\right) -\frac{1}{2} \xi^2
\left(\frac{1}{2}\gamma^{kl}{\mathcal{B}}^2-{\mathcal{B}}^k{\mathcal{B}}^l\right)
~.
\end{equation}
Now in making our conformal transformation we recognise that our space
metric will tend to the Schwarzschild form at infinity with $m=GM$
being the Schwarzschild asymptotic mass parameter. Since Schwarzschild space is
conformally 3-flat we may impose the boundary condition $\Phi
\rightarrow O(1/r)$ on the solution $\Phi$ of
(\ref{eq:20}). Then $\Phi$ tends to the corresponding Schwarzschild
value $m/r$ found in equation (\ref{eq:8}) of the introduction as
$r\rightarrow\infty$, that is we identify the coefficient of $1/r$ as
the total mass parameter. When we integrate equation (\ref{eq:24}) over all
space so as to generate the mechanical energy $E_M$ from the
first term on the right, we find on the left
$$
\int \nabla\cdot \left(\frac{1}{2}\xi^2
{\bf{\mathcal{A}}}\times{\bf{\mathcal{B}}}
-2{\bf{\nabla}}\Phi\right)dV =
\int\left(\frac{1}{2}\xi^2{\bf{\mathcal{A}}}\times{\bf{\mathcal{B}}}
-2{\bf{\nabla}}\Phi\right)\cdot d{\bf S},
$$
where the $d{\bf S}$ integral is to be evaluated over the sphere at
infinity still assuming no black hole is present. Now ${\bf{\mathcal{B}}}$ is $O(1/r^3)$
there, the triad component of ${\bf{\mathcal{A}}}$ will be
$O(1/r^2)$ like the classical vector potential and
$\xi^2\rightarrow 1$ so the integral of the first term on the right
vanishes; however $\Phi\rightarrow  m/r$, so the second term
gives $8\pi m=\kappa M$. Using this result, the integration of
(\ref{eq:24}) over all space yields
\begin{equation} \label{eq:29}
E_G=M-E_M=\int\rho_G dV~,
\end{equation}
with $\rho_G$ given in (\ref{eq:25}).

As KLB [6] pointed out the mechanical and gravitational energy densities
defined above fail for systems with ergospheres because they involve
$w^\mu=\xi^\mu/\xi$ in their definition and $\xi$ is zero on the
ergosphere. However such difficulties can be circumvented by using the
mechanical energy as estimated by hypersurface orthogonal observers
such as Bardeen's ZAMOs. We take such observers moving orthogonally to
surfaces of constant time with 4-velocities
$\tilde{w}^\mu=\zeta^{-1}(1,W^k)$. The appropriate form of metric
is in terms of lapse and shift written as the final expression of
(\ref{eq:12}). The metric components and their inverses were given in
equations 2.37 and 2.38 of KLB [6] and we notice that $W_k=\xi_k$:
\begin{eqnarray} \label{eq:30}
g_{00}&=&\zeta^2-W^2~~,~~g_{0l}=W_l=\tilde{\gamma}_{kl}W^k~~,~~g_{kl}=-\tilde{\gamma}_{kl}~,\nonumber\\
g^{00}&=&\zeta^{-2}~~,~~g^{0l}=\zeta^{-2}W^l~~,~~g^{kl}=-\tilde{\gamma}^{kl}+\zeta^{-2}W^kW^l~,\nonumber\\
\sqrt{-g}&=&\zeta\sqrt{\tilde\gamma}~~~,~~~\tilde\gamma={\rm det}\tilde\gamma_{kl}.
\end{eqnarray}
The Einstein equations are
\begin{equation} \label{eq:31}
\kappa T^{00}=G^{00}=\frac{1}{2}\zeta^{-2}(\tilde{P}+K^2-K^{kl}K_{kl})~,
\end{equation}
\begin{equation} \label{eq:32}
\kappa T^0_k =G^0_k=-\zeta^{-1}\tilde{\nabla}_l(K^l_k -\delta^l_K K)~,
\end{equation}
\begin{equation} \label{eq:33}
-\kappa T=R=2(\zeta^2 G^{00} -\tilde{\gamma}^{kl}R_{kl})~,
\end{equation}
\begin{equation} \label{eq:34}
\kappa
T_{kl}=R_{kl}=\tilde{P}_{kl}+\tilde{\nabla}_{(k}\tilde{\mathcal{E}}_{l)}
-\tilde{\mathcal{E}}_k\tilde{\mathcal{E}}_l+2\zeta
K^m_{(l} \nabla_{[l)}W_{m]} -\zeta^{-1}\tilde{\mathcal{E}}_m W^m
K_{kl}+\nabla_m (W^m K_kl)~,
\end{equation}
where $\tilde{P}_{kl}$ is the Ricci tensor of the spatial metric
$\tilde{\gamma}_{kl},~\tilde{P}=\tilde{\gamma}^{kl}\tilde{P}_{kl}$ is
the corresponding 3-scalar curvature and $K_{kl}$ is the second
fundamental form of the hypersurface $t=$constant. 
Thus
\begin{equation} \label{eq:35}
K_{kl}=\zeta^{-1}\tilde{\nabla}_{(k} W_{l)}~~,~K=K^k_k
=\tilde{\gamma}^{kl}K_{kl}~~,~~{\rm and}~~
\tilde{\mathcal{E}}_k=-\partial_k\ln\zeta~.
\end{equation}

The mechanical energy density on any hypersurface $\Sigma$ of normal
$n^\mu$ is for dust $\rho({\bf u.n})^2$ so for our observers moving
orthogonally to the cut $n^\mu=\tilde{w}^\mu$ and for general
$T^{\mu\nu}$ the mechanical energy on an element of hypersurface
$\sqrt{-g}d\Sigma_\mu$ is $T^{\mu\nu}\tilde{w}_\nu\sqrt{-g}d\Sigma_\mu
=\zeta^2 T^{00}\sqrt{\tilde{\gamma}}d^3x$. This is given directly by
equation (\ref{eq:31}). Once again $\tilde{P}$ is determined by using
a conformal transformation $\tilde{\Phi}$ such that the transformed
$\tilde{\gamma}_{kl}$ has vanishing 3-curvature. Analogously to (\ref{eq:20}) this
leads to 
\begin{equation} \label{eq:36}
-2\tilde{\nabla}^2\tilde{\Phi}+|\tilde{\nabla}\tilde{\Phi}|^2=
 \frac{1}{2}\tilde{P}~.
\end{equation}
Transporting second derivative terms to the left this yields in place
of (\ref{eq:31})
\begin{equation} \label{eq:37}
\kappa \zeta^2 T^{00}+2\tilde{\nabla}^2\tilde{\Phi}
=|\tilde{\nabla}\tilde{\Phi}|^2 +\frac{1}{2}(K^2-K_{kl}K^{kl})~.
\end{equation}
Integrating and using the boundary condition that
$\nabla\tilde{\Phi}\rightarrow (GM/r^2) \hat{\bf r}$ at infinity
we find  for maximal slices $(K=0)$
\begin{equation} \label{eq:38}
-\tilde{E}_G=\tilde{E}_M -M=\frac{1}{\kappa} \int
 \left[|\tilde{\nabla}\tilde{\Phi}|^2 -\frac{1}{2}
 K_{kl}K^{kl}\right] dV~.
\end{equation}

The mechanical energy density $\tilde{E}_M$ has real advantages over
$E_M$. Not only can it be measured within the ergosphere but also the
two `$\gamma$' factors $({\bf u.n})$ are the same and correspond to
what a hypersurface orthogonal observer would see. It also coincides with the geometrical expression of Wald alluded to earlier. However there are
strong arguments against it also. The observers move relative to
static observers and worse still they move relative to each other. As
seen from infinity they circulate and for the axially symmetric case
with azimuthal Killing vector $\eta^\mu$ they rotate about the axis at
angular velocity $\omega={\boldsymbol \xi.\boldsymbol\eta/(-\boldsymbol\eta.\boldsymbol\eta)}$ which depends on
position. In this case the space-time cuts correspond to surfaces of
constant Boyer-Lindquist time. Now in classical physics observers who
rotate at angular velocity ${\bf \Omega}$ see as their energy not the
true energy $E$ relative to observers at rest at infinity but rather
the Jacobi constant $E-{\bf \Omega.J}$, where ${\bf J}$ is the total
angular momentum. For observers such as ZAMOs in differential rotation
we expect $E-\int \omega jd\tilde{V}$, where $j$ is the angular
momentum density about the symmetry axis. It is interesting that our
expression for $\tilde{E}_M$ takes just such a form (writing $d\tilde V=\sqrt{\tilde\gamma}d^3x$): 
\begin{equation} \label{eq:39}
\tilde{E}_M=\int T^{\mu\nu} \tilde w_\nu\sqrt{-g}d\Sigma_\mu =\int
T^0_\nu \zeta^\nu d\tilde V=\int T^0_\nu(\xi^\nu
+\omega\eta^\nu) d\tilde V 
={\mathcal T}-\int \omega j d\tilde V,
\end{equation}
where ${\mathcal T}=\int T^0_0\sqrt{\tilde{\gamma}}d^3x$ is the part of the
mechanical energy independent of the angular velocity of the observer
and $j=-T^0_\nu\eta^\nu$ is the angular momentum density. If, following this
line of thought, we identify ${\mathcal T}$ with the mechanical energy rather
than $\tilde{E}_M$ then we need to evaluate $T^0_0$ rather than
$T^{00}$. Einstein's equations in mixed form follow from (\ref{eq:31})
and (\ref{eq:32}) by writing $G^0_0=(g^{00})^{-1}(G^{00}-g^{0k}G^0_k)$;
hence using (\ref{eq:36}) and putting all second derivatives on the
left
\begin{equation} \label{eq:41}
\kappa T^0_0 +2\nabla^2\tilde{\Phi} -\nabla_l
[\zeta^{-1}W_k(K^{kl}-\tilde{\gamma}^{kl}K)]
=|\tilde{\nabla}\tilde{\Phi}|^2 +\frac{3}{2}(K^2-K^{kl}K_{kl})
-\zeta^{-1}W^k (K^l_k -\delta^l_k   K)\tilde{\mathcal E}_l~.
\end{equation}
On integration the third term vanishes over the sphere at infinity and in the absence of black holes
we have, setting $K=0$ for a maximal $t=$constant slice,
\begin{equation} \label{eq:42}
-{\mathcal{V}}={\mathcal T} -M= \frac{1}{\kappa}\int [|\nabla\tilde{\Phi}|^2
 -\frac{3}{2}
K^{kl}K_{kl}-\zeta^{-1}W_kK^{kl}\tilde{\mathcal {E}}_l]d\tilde{V}, 
\end{equation}
which expresses the new gravitational potential energy ${\mathcal{V}}$
in terms of `field' quantities.

Advantages of $\mathcal {T}$ as a mechanical energy are: 

(i) like $\tilde{E}_M$ it can be evaluated for systems with
ergospheres;

(ii) like $E_M$ it involves $m{\bf u\cdot{\boldsymbol \xi}}$ which is the energy for a
dust particle, even for one that darts into and out from the
ergosphere.

(iii) It has removed the part of $\tilde{E}_M$ which is related to the
circulation of the ZAMOs around the axis.

However, even for dust it is not clear that it can be related to
kinetic energy as seen by any chosen observers.

\section{Mechanical Angular Momentum Densities}

For axially symmetrical systems there is no difficulty in defining
total angular momentum. It is   $-\int T^\mu_\nu \eta^\nu \sqrt{-g}~d\Sigma_\mu$ and this
translates into $-\int T^\mu_\nu \eta^\nu\zeta d\tilde{V}$, or, for
stationary systems without ergospheres $-\int T^\mu_\nu \eta^\nu\xi
dV$. However, whereas the split between
mechanical energy and gravitational energy was clear, the split
between mechanical angular momentum and field angular momentum caused
us difficulty. Luckily considerable enlightenment comes from first
studying the electrodynamic analogue in flat space, so to this we now
turn. We consider a charged rotating fluid held together by some
cohesive force such as surface tension; a charged oil drop might be a
good example. We wish to split the total angular momentum into a part
due to the mechanics of the fluid itself and an electromagnetic
part. We start by considering a single particle of mass $m$ and charge
$q$. Its 3-velocity will be ${\bf v}$ or in cylindrical polar
coordinates $(\dot{R}, R\dot{\phi},\dot{z})$, but that is written in
local triad language; a relativist would use the 3-metric
$d\sigma^2=dR^2+\gamma_{\phi\phi} d\phi^2+dz^2$ and write the
3-velocity as $v^k=(\dot{R},\dot{\phi},\dot{z})$ or
$v_k=(\dot{R},\gamma_{\phi\phi}\dot{\phi},\dot{z})$, where in this
case $\gamma_{\phi\phi}$ is just $R^2$. As recorded by static
observers the mechanical angular momentum of a particle is just
$m\gamma_{\phi\phi} (d\phi/d\tau)$, where
$d\tau=dt\sqrt{1-v^2}$; however this is not the total angular momentum
associated with the particle because it is charged and moves in an
electromagnetic field. Its Lagrangian will be $-m\sqrt{1-v^2} +q{\bf
v}.{\bf A}-qA_0$ where ${\bf A}$ is the electromagnetic vector
potential and $A_0$ is the electrostatic potential. Thus $\partial
L/\partial\dot{\phi}=p_\phi$ will have two pieces:
$ m\gamma_{\phi\phi}\dot{\phi}/\sqrt{1-v^2} +qA_\phi$. The
first is the mechanical angular momentum
$m\gamma_{\phi\phi}(d\phi/ d\tau)$, considered above; the second
is a piece of electromagnetic angular momentum that is to some extent
associated with the motion of the particle and its charge, but it also
depends on all the other charges that act together to give rise to the
vector potential ${\bf A}$. Notice this piece of momentum is not
generally gauge invariant because we have not yet said anything about
fixing the electromagnetic gauge. It is actually incorrect to believe
that the sum over all particles of all their $p_\phi$ gives the total
angular momentum. It does not! The trouble arises just because the
$A_\phi$ is generated by other particles in the same assemblage --- we
encounter a similar problem when adding individual particle energies
in which the electrical potential is included. Even classically the
straight sum counts the electrostatic energy twice. What then is the
correct procedure in the electrodynamic case? The answer is to
separate the mechanical angular momentum which is additive and gauge
invariant. We sum this over all the particles. Then quite separately
we work out the total electromagnetic angular momentum from the gauge
invariant expression $J_{em}=\frac{1}{4\pi}\int[{\bf r} \times ({\bf
E} \times {\bf H})].\hat{\bf{z}} d^3x = \int M^0_\phi \eta^\phi d^3x$,
where $M^\mu_\nu$ is the Maxwell stress-energy-momentum of the
electromagnetic field and $\eta^\mu$ is the angular Killing vector, so
$\eta^\phi=1$. When we sum $J_M+J_{em}$ we find that we do indeed get
the total angular momentum. This is obvious since the total
$T^{\mu\nu}$ is the sum of the mechanical and the electromagnetic
energy momentum tensors. This final more obvious option is not open to
us when we deal with the gravitational field's angular momentum, as
there is no known stress energy tensor for it; however, as we now
show, the earlier argument for splitting the mechanical angular
momentum from the field part can be followed in the gravitational case
to which we now return.

We shall again consider a single particle but now it will be uncharged
and in a stationary metric as in  (\ref{eq:12}), i.e.,
\begin{equation} \label{eq:A1}
ds^2=\xi^2(dt-{\mathcal{A}}_kdx^k)^2
-\gamma_{kl} dx^k dx^l~,
\end{equation}
but we shall take this metric to be axially symmetric and, to start with, we
consider the still simpler case in which there is a $\phi\rightarrow
-\phi,~t \rightarrow -t$ symmetry. In the latter case we write
${\mathcal{A}}_kdx^k={\mathcal{A}}_\phi d\phi +{\mathcal{A}}_Ldx^L$ and 
\begin{equation} \label{eq:A2}
\gamma_{kl}dx^kdx^l=\gamma_{\phi\phi} d\phi^2
+\gamma_{KL}dx^Kdx^L~,
\end{equation}
where $K$ and $L$ run from 2 to 3 whereas $k,l$ run from 1 to 3 and
Greek suffices run from 0 to 3.  

\section{Mechanical Angular Momentum from Static Observers}

To keep contact with directly observable quantities we introduce our
set of static observers (see Section II). The speed of the fluid, $v$, past the
observers is given as $(1-v^2)^{-1/2}=w^\mu u_\mu =({\bf w.u})$ and
the components of $\frac{\bf v}{\sqrt{1-v^2}}$ are given by ${\bf
u-(u.w)w}$. Hence the mechanical momentum of the particle, unconnected
with its field momentum, will be $m[{\bf u-(u.w)w}]$ and the
corresponding mechanical angular momentum is $m({\bf
u-(u.w)w}).\boldsymbol{\eta}$, $\boldsymbol\eta$ is the axial Killing vector. Since only the component of
$\boldsymbol{\eta}$ transverse to $\xi^\mu$ is involved in this
product, this angular momentum can also be re-expressed as
  $m~
u^\mu \gamma_{\mu\nu}\eta^\nu$, where $\gamma_{\mu\nu}=w_\mu
w_\nu-g_{\mu\nu}$. If we consider a stress-energy tensor made up of
dust, then $T^{\mu\nu}=\rho u^\mu u^\nu$ and $\rho u^\mu$ is the rest
mass flux vector; multiply by $u^\nu \gamma_{\nu\sigma} \eta^\sigma$,
the mechanical angular momentum per unit rest mass, and $\rho u^\mu
u^\nu\gamma_{\nu\sigma}\eta^\sigma=T^{\mu\nu}\gamma_{\nu\sigma}\eta^\sigma$
is the flux of mechanical angular momentum. We shall now find this by
another method directly analogous to our electrodynamic calculation.

The Lagrangian for a free particle moving in the metric (\ref{eq:A1})
is $L=-m[\xi^2(1-{\mathcal{A}}_\phi \dot{\phi} -{\mathcal{A}}_K
\dot x^K)^2 -\gamma_{\phi\phi}\dot{\phi}^2 -\gamma_{KL}\dot{x}^K
\dot{x}^L]^{1/2} =-m~ds/dt$. Its total angular momentum will be
$p_\phi =\partial L/\partial\dot{\phi}$, so 
\begin{eqnarray*}
p_\phi &=& m[\xi^2(1-{\mathcal{A}}_\phi\dot{\phi}-{\mathcal{A}}_K
\dot{x}^K){\mathcal{A}}_\phi
+\gamma_{\phi\phi}\dot{\phi}]/(ds/dt)~,\\
&=& m\xi^2(u^0-{\mathcal{A}}_\phi u^\phi -{\mathcal{A}}_K
u^K) {\mathcal{A}}_\phi +m\gamma_{\phi\phi}u^\phi~.
\end{eqnarray*}
Now $u^\mu=(dt/ds,~dx^k/ds)$,  so $ u_0=\xi^2(u^0-{\mathcal{A}}_\phi
u^\phi-{\mathcal{A}}_K u^K)$ and $(1-v^2)^{-1/2}=w^\mu
u_\mu=\xi^{-1}u_0$, so reversing the order of the terms we get 
$p_\phi=m\gamma_{\phi\phi}
u^{\phi}+m\xi(1-v^2)^{-1/2}{\mathcal{A}}_\phi$.
This expression is in precise correspondence with the electromagnetic
case with the mechanical angular momentum in the first term and the
gauge dependent field term coming afterwards. When the system does not
have $\phi\rightarrow-\phi$, $t\rightarrow -t$ as a symmetry, there are
$d\phi dx^K$ terms in $\gamma_{kl}dx^k dx^l$, and the form of
$p_\phi/m$ becomes
$$p_\phi/m =\gamma_{\phi k}u^k +\xi(1-v^2)^{-1/2}{\mathcal{A}}_\phi~.$$
The mechanical angular momentum per unit mass is the first term and to
get the angular momentum flux vector we multiply this by the rest mass
flux $\rho u^\mu$, so the mechanical angular momentum flux remains $T^{\mu
k}\gamma_{k\phi}=T^{\mu\nu}\gamma_{\nu\phi} = T^{\mu\nu}
\gamma_{\nu\sigma}\eta^\sigma$. 

Although for easy explanation we have adopted the simplest dust case
to explain our points, nevertheless our final formulae still hold
whatever the constitution of $T^{\mu\nu}$.

The total gravitational field angular momentum reckoned by static
observers is the difference between the total angular momentum and the
total mechanical angular momentum, so for a spacelike surface $\Sigma$
$$J_{GS}=J-J_{MS}= \int -T^{\mu\nu}(g_{\nu\sigma} +\gamma_{\nu\sigma})
\eta^\sigma \sqrt{-g}d\Sigma_\mu =\int -T^{\mu\nu}w_\nu w_\sigma
\eta^\sigma \sqrt{-g}d\Sigma_\mu =-\int T^0_0\xi{\mathcal{A}}_\phi
dV~.$$
Now we have established this as the angular momentum they
attribute to the gravitational field our next aim is to re-express it
in terms of the squares of field quantities by using Einstein's
equations for $T_{\mu\nu}$ and performing integrations by
parts. Multiplying (\ref{eq:24}) by $\xi{\mathcal{A}}_\phi$, integrating
over all space we find on smuggling the $\xi{\mathcal{A}}_\phi$ into
the divergence and then paying the duty 
$$
\int\xi{\mathcal{A}}_\phi{\bf{\mathcal{D}}}\cdot d{\bf S}
-\int{\bf{\nabla}}(\xi{\mathcal{A}}_\phi)\cdot {\bf{\mathcal{D}}}dV=
\kappa\left(\int T^0_0\xi{\mathcal{A}}_\phi dV+\int
\rho_G\xi{\mathcal{A}}_\phi dV\right).
$$
Now ${\mathcal{A}}_\phi$ is $O(1/r)$ and ${\mathcal{D}}$
is $O(1/r^2)$, so the first term vanishes when
integrated over the sphere at infinity while the first term on the
right gives $-J_{GS}$. Hence,   
$$
J_{GS} =\int\left[\rho_G\xi{\mathcal{A}}_\phi +\frac{1}{\kappa}
{\bf{\nabla}} (\xi{\mathcal{A}}_\phi)\cdot {\bf{\mathcal{D}}}\right] dV,
$$
where ${\mathcal{D}}$ is given in (\ref{eq:24}) and $\rho_G$ in  (\ref{eq:25}). When
${\bf{\mathcal{A}}}$ has only a $\phi$ component,
${\bf{\nabla}}_k{\mathcal{A}}_\phi=-\eta_{\phi km}{\mathcal{B}}^m$, so
then ${\bf{\mathcal{D}}}\cdot {\bf{\nabla}}\phi=-({\mathcal{D}} \times
{\mathcal{B}})_\phi$ and the final term becomes
$\frac{\xi}{\kappa}[-({\mathcal{D}} \times {\mathcal{B}})_\phi
-{\bf{\mathcal{D}}}\cdot {{\mathcal{E}}}\mathcal{A}_\phi]$. 

\section{Mechanical Angular Momentum with respect to ZAMO's}

The ZAMOs have a different concept of what constitutes space than the
static observers and that concept is superior in that it extends
within ergospheres. It is natural to replace the $w_\nu$ of the static
observers in equation (\ref{eq:14}) by the $\tilde{w}_\nu=n_\nu$ of
the ZAMOs. Indeed if we calculate $p_\phi$ in the appropriate
ZAMOs coordinates, we find
$$
p_\phi=\frac{\partial L}{\partial\dot{\phi}} =m\tilde{\gamma}_{\phi
k}(u^kW^ku^0),
$$
and it is just the $u^k$ term that the ZAMOs consider to be the
mechanical angular momentum with respect to them. Their estimate of
mechanical angular momentum is therefore (with $\phi$ unsummed)
$$
\tilde{J}_M=\int T^{\mu\nu}\tilde{\gamma}_{\nu\sigma}\eta^\sigma
\sqrt{-g}d\Sigma _\mu~.
$$
 But $J=-\int
T^{\mu\nu}g_{\nu\sigma}\eta^\sigma \sqrt{-g}d\Sigma_\mu$ and
$\eta^\sigma$ has only a $\phi$ component and since
$\tilde{\gamma}_{\mu\nu}=-g_{\mu\nu} +\tilde{w}_\mu \tilde{w}_\nu$, it
follows that $\tilde{J}_M-J=\int
T^{\mu\nu}\tilde{w}_\nu\tilde{w}_\sigma \eta^\sigma
\sqrt{-g}~d\Sigma_\mu=0$ because $\tilde{w}_\sigma \eta^\sigma =0$.

Thus the total angular momentum
is the same as the material angular momentum as seen in the local rest
frames of the ZAMOs.

The mechanical angular momentum defined via static observers cannot be
assessed for systems with ergospheres. The mechanical angular momentum
defined via ZAMOs takes no account of their circulation around the
axis. Is there a splitting into mechanical and field angular momentum
that takes account of the ZAMOs motion but does not break down within
ergospheres? Something like the split we found for energy by writing
$\zeta^\mu =\xi^\mu+\omega \eta^\mu$ is needed.

For a dust fluid the angular momentum is $J=-\int\rho u^\mu u_\nu
\eta^\nu \sqrt{-g}~d\Sigma_\mu =-\int \rho u^0 u^\nu g_{\nu\phi}
d\tilde{V}$.
Now $u^\nu g_{\nu\phi} = u^0 g_{0\phi} +u^k g_{k\phi} =u^0W_\phi
-u^k \tilde{\gamma}_{k\phi}$.
The first term on the right is due to the circulation of the ZAMOs,  the
last is most readily seen as an angular momentum per unit mass of the
fluid when $u^k$ has only a $\phi$-component.
Thus,  following Bardeen (1970) [7],  $J$ can be split as follows
$$
J=-\int T^{00}W_\phi\zeta d\tilde{V} +\int
T^{0k}\tilde{\gamma}_{k\phi}\zeta d\tilde{V}~.
$$
For those who want this split expressed covariantly the first `field'
term may be rewritten
$$
J_f=-\int T^{\mu\nu}\tilde{w}_\nu\zeta^{-1}(\xi
.\eta)\sqrt{-g}~d\Sigma_\mu~.
$$

To evaluate the field angular momentum we rewrite (36) in a form
suitable for constructing the required integral (setting $K^2=0$ for a
maximal slice):
\begin{equation} \label{eq:44}
\kappa T^{00}\zeta W_\phi+2 \tilde{\nabla} \cdot (\zeta^{-1}W_\phi
\tilde{\nabla}\tilde{\Phi}) =
\zeta^{-1}W_\phi|\tilde{\nabla}\tilde{\Phi}|^2 +2
\tilde{\bf{\nabla}}(\zeta^{-1}
W_\phi)\cdot \tilde{\bf{\nabla}}\tilde{\Phi}-\frac{1}{2} \zeta^{-1} W_\phi
K_{kl}K^{kl} =-\kappa j_f~.
\end{equation}
Integrating,  we have (after division by $-\kappa$)
$$
J_f=\int j_f d\tilde{V},
$$
where $j_f$ is defined in equation (\ref{eq:44}). $\zeta$ and $W_\phi$ can be expressed in terms of the Killing vectors $\xi$  and the angular one $\eta$ via $W_\phi=\xi\cdot \eta$ and $\zeta^2=\xi^2- (W_\phi^2/\eta^2)$. 

\section{Conclusions}

We have elucidated the physics of three different expressions for the
energy and angular momentum densities of gravitational fields. While
the authors differ as to the relative importance of these expressions
DL-B favours the third.

None of these densities make contributions to $T^{\mu\nu}$ although they
do act as (negative) sources for the conformal field $\Phi$ which
gives the scalar curvature of 3-space.

The fact that they do not contribute to the source $T^{\mu\nu}$ was
one of MTW's strongest arguments for the non-existence of a
stress-energy tensor for gravity. That argument still stands. We have
conducted our investigation within General Relativity. A different
theory might have the field energy contributing to the source
$T^{\mu\nu}$ but that is not General Relativity. Energy densities are
only really useful when the system changes, whereas this paper is
confined to stationary systems. Nevertheless, we believe these methods
may be generalised to systems that change so slowly that no
gravitational waves are emitted.

 \Large{\bf Acknowledgements}
 \vskip .5 cm
 \normalsize 
 \setlength{\baselineskip}{20pt plus2pt}
 We acknowledge the support of the grant GA\v CR 202/06/0041 of the Czech Republic, the Royal Society Grant (JB, DLB),  and the hospitality of the Institute of Astronomy, Cambridge (JB, JK).
 
 JB also acknowledges the support of the grants LC06014 and MSM0021620860 of the Ministry of Education and Humboldt Award.

{}

\end{document}